\documentclass[twocolumn,twoside,slac_two]{revtex4}
\usepackage{graphicx}
\usepackage{fancyhdr}
\begin{document}

\title{Multi-wavelength emission region of $\gamma$-ray pulsars}

%

\author{S. Kisaka, Y. Kojima}
\affiliation{Department of Physics, Hiroshima University, Higashi-Hiroshima 739-8526, Japan}
%

\begin{abstract}

Recent obserbations by {\it Fermi Gamma-Ray Space Telescope} of $\gamma$-ray pulsars have revealed further details of the structure of the emission region. We investigate the emission region for the multi-wavelength light curve using outer gap model. We assume that $\gamma$-ray and non-thermal X-ray photons are emitted from a particle acceleration region in the outer magnetosphere, and UV/optical photons originate above that region. We also assume that $\gamma$-rays are radiated only by outwardly moving particles, whereas the other photons are produced by particles moving inward and outward. We parametrize the altitude of the emission region. We find that the outer gap model can explain the multi-wavelength pulse behavior. From observational fitting, we also find a general tendency for the altitude of the $\gamma$-ray emission region to depend on the inclination angle. In particular, the emission region for low inclination angle is required to be located in very low altitude, which corresponds to the inner region within the last-open field line of rotating dipole in vacuum. This model suggests a modification of statistics about observed $\gamma$-ray pulsars.

\end{abstract}

\maketitle

\thispagestyle{fancy}

\section{Introduction}
  Pulsars emit over a wide range of energies from radio to $\gamma$-ray. Recent observations by the {\it Fermi Gamma-Ray Space Telescope} of more than eighty pulsars \citep{Ab11} have revealed further details of the structure of the emission region. The light curve in the $\gamma$-ray band is an important tool for probing the particle acceleration and dissipation processes in the pulsar magnetosphere. The $\gamma$-ray emission region has therefore been explored by comparing theoretical models with the observed light curve(e.g., \citep{Wa09, VHG09, RW10}). The pulsed emission is also detected in other energy bands (X-ray, ultraviolet, optical and radio) for some sources. The spectral features are non-thermal except for the soft X-ray range, and the light curves from a single object are, in general, different from one energy band to another. A complete understanding of light curve behavior in multi-wavelength bands can provide valuable information about the particle acceleration region. 

 Possible origins of non-thermal pulsed emissions have been considered in the polar cap \citep{DH96}, slot gap \citep{MH04}, and outer gap \citep{CHR86} models. Recent {\it Fermi} observations \citep{Ab09} rules out the near-surface emission proposed in polar cap cascade models, which would exhibit a much sharp spectral cutoff due to magnetic pair-production attenuation. Thus, pulsed $\gamma$-ray emission originates in the outer magnetosphere, as considered in the outer gap model.

Here, we investigate the emission regions of several pulsars by fitting the simplified model of Takata et al. (2008, hereafter TCS08) \citep{TCS08} to the observed multi-wavelength light curves. In this model, we have to specify the locations of the upper and lower boundaries of the gap region where the non-corotation potential is zero. Therefore, we explicitly introduce the altitude of the gap region as a parameter, in order to fit the observational data easily. The light curves also depend on the dipole inclination and viewing angles. In our method, such parameters are eliminated by other observational data, and only the altitude is changed for the fitting. In the most studies, the lower boundary of the emission region is chosen as the surface of the last-open field lines of the rotating dipole (e.g., \citep{Wa09}). In this study, however, the altitude is allowed to be in a wide range in order to explore the possible deviation of magnetic field-line structure from that of a rotating dipole in vacuum.

\section{Model}

The numerical method for fitting the light curve is well described by Romani \& Watters (2010) \citep{RW10} and Bai \& Spitkovsky (2010a) \citep{BS10}. We follow this method except one modification that we explicitly introduce the altitude of the emission region as an additional parameter. We assume that magnetic field structure is approximately described by a rotating dipole and that radiation direction aligns with magnetic field in a frame rotating with angular velocity in which the electric field vanishes.

A certain mechanism is needed to fix the lower boundary of the particle acceleration region. In most works, including TCS08, the lower boundary is chosen as the surface of the last-open field lines of a rotating dipole in a vacuum. In the outer gap model, if particle acceleration occurs in an open zone, the curvature radiation from the accelerated particles forms a narrow cone along the magnetic field lines in a frame rotating with angular velocity. These $\gamma$-ray photons are converted by colliding X-ray photons to $e^{\pm}$ pairs, which tend to screen the accelerating electric field. However, there is no supply of pairs on the last-open field lines and hence no screening of the electric field, since the $\gamma$-ray photons are emitted only toward higher altitudes above the last-open field lines \citep{CHR86}. The `real' last-open field lines may be different from ones in a vacuum. We therefore take into account this possible deviation of the boundary. We assume that dipole magnetic field is an approximation within the light cylinder and use rotating diople field as the global magnetic field structure. Even if the overall structure is not different so much, critical value between open and closed field lines is very sensitive to the boundary value  at the surface. Thus we introduce a parameter, altitude of the emission region as a  correction factor in order to take into account the deviation of boundary from the vacuum field. In our model this parameter specifies the range of the emission region which is located above or below the last-open field lines within the light cylinder radius. Each different field line originating from the magnetic polar region is parameterized by magnetic colatitudes $\theta_{\rm m}$ and azimuthal angles $\phi_{\rm m}$. Following Cheng et al. (2000) \citep{CRZ00}, we define open volume coordinates on the polar cap, ($r_{\rm ov}$, $\phi_{\rm m}$), where $r_{\rm ov}\equiv \theta_{\rm m}/\theta^{{\rm pc}, 0}_{\rm m}(\phi_{\rm m})$. The function $\theta^{{\rm pc}, 0}_{\rm m}$ is the magnetic colatitude of the conventional polar cap angle and generally depends on the magnetic azimuth $\phi_{\rm m}$. The parameter $r_{\rm ov}$ corresponds to the altitude of the emission region: The last-open field lines of a rotating dipole in a vacuum correspond to $r_{\rm ov}=1$, whereas those for higher altitudes have $r_{\rm ov}<1$. The maximum value is chosen as $r_{\rm ov}= 1.36^{1/2}$, which corresponds to the polar cap angle $\theta^{\rm pc}_{\rm m}\sim 1.36^{1/2}\theta^{{\rm pc}, 0}_{\rm m}$, obtained in the force-free limit by Contopoulos et al. (1999) \citep{CKF99}. 

We assume that the radiation of different energy bands is emitted from different field lines characterized by altitude. The field line relevant to the $\gamma$-ray and X-ray is approximated as being the same one. The direction of the emission is tangential to the lines, and inward and outward directions are possible. Both location and direction affect the light curve profile of the energy bands. Following the model by TCS08, the $\gamma$-ray radiation above 100 MeV is emitted by particles moving in an outward direction, whereas radiation at lower energy bands is emitted by those moving in both outward and inward directions. We use two conditions to constrain the emission region. First condition is the radial extension of the emission region. The outward emission is restricted to radial distances $r_{\rm n} < R_{\rm LC}$, and the inward one is restricted to $r_{\rm s} < r < \min (3r_{\rm n}, R_{\rm LC})$. The outer boundary $3r_{\rm n}$ for inward emission comes from the results of dynamic model (TCS08), in which very few ingoing pairs are produced beyond the radial distance $r > 3r_{\rm n}$. Second condition is the azimuthal extension of the emission region. We use the magnetic azimuthal angle of the footprint of field line (i.e., the point where magnetic field line penetrates the neutron star surface) to characterize the field line for given $r_{\rm ov}$. Radial distance to the null charge surface on the field lines significantly depends on the magnetic azimuthal angle. In the outer gap model, most of the pairs are created around the null surface (TCS08). We expect that the gap activity is related to the distance to the null surface. Although the current density should be determined by global conditions, there is no study of the three-dimensional magnetosphere of an inclined rotator. Here, we assume that the field lines of both outward and inward emission are active only if the radial distance to null surface $r_{\rm n}$ is shorter than $R_{\rm LC}$. Spatial distribution of the emissivity is approximated by the step function-type, but the peak positions weakly depend on the detailed emissivity distribution.

We assume that the overall structure of the light curve comes not from the emissivity distribution,  but from a bunch of many field lines in the observation, that is, caustics. Here, we focus on the peak phases of the light curve, so we adopt a simple, uniform emissivity along all magnetic field lines, which is independent of both the magnetic azimuthal angle $\phi_{\rm m}$ and the altitude $r_{\rm ov}$.

In the observed light curve, the reference phase $\phi=0$ is assumed to be located at the radio emission peak maximum. However in the model light curve, the conventional reference phase $\phi=0$ occurs when the magnetic axis, spin axis and Earth line of sight lie all in the same plane. These two reference phases do not agree with each other since it is generally assumed that radio emissions arise at non-zero altitude in most empirical studies. Following Romani \& Watters (2010) \citep{RW10}, we allow a shift by $-0.1 \le \delta\phi \le 0.1$ in the model reference rotation phase. This degree of freedom does not significantly affect the determination of the altitude parameter $r_{\rm ov}$. For more details on this model see Kisaka \& Kojima (2011) \citep{KK11}.

\section{Results}

Here, we compare our model with pulse profiles observed at multiple wavelengths for seven pulsars. The sources are chosen using two criteria. One is that non-thermal pulses are detected in addition to the $\gamma$-ray and radio bands. Our concern is to explore whether or not the emission region for different energy bands is explained by the TCS08 model. The second criterion is that the geometrical parameters, inclination angle $\alpha$ and viewing angle $\xi$ are observationally constrained by the relativistic Doppler-boosted X-ray pulsar wind nebula or radio polarization data. The torus fitting method constrains the viewing angle $\xi$ only. A small allowed range of $|\alpha-\xi|\leq 10^{\circ}$  is assumed for samples in which only $\xi$ is constrained due to the fact that radio emission from the pulsar polar region is detected. The geometrical parameters for our selected pulsars are listed in Table 1. We use these values, although there are some uncertainties in them. The results are summarized in Fig. 1, which shows the intensity map for outward (upper panel) and inward (lower panel) emission as a function of the altitude of the emission region $r_{\rm ov}$ and rotational phase. This shows that the peak phases of seven pulsars emitting $\gamma$-ray and X-ray can be successfully fitted using the TCS08 outer gap model, in which both $\gamma$-rays and X-rays originate from the same magnetic field line characterized by an altitude $r_{\rm ov}$. The parameter $r_{\rm ov}>1$ is needed in the light curve fitting for some sources. Moreover, the inclusion of inward emission for X-rays causes a variety of pulse profiles in both bands. The parameter $r_{\rm ov}$ could not be determined solely using $\gamma$-ray data for a single $\gamma$-ray peak pulsar. However, by considering the X-ray light curve, the parameter is uniquely determined for PSRs J0659+1414, J2229+6114 and J1420-6048. For details of each pulsars, see Kisaka \& Kojima (2011) \citep{KK11}.

\begin{table}
\begin{tabular}{ccc}
\multicolumn{3}{c}{TABLE 1 Geometrtical parameters} \\ \hline
Name & $\alpha$ & $\xi$   \\  
  &(degrees) &(degrees)   \\
(1) & (2) & (3)  \\ \hline
J0835-4510 & 72 & 64 \\
J0659+1414 & 29 & 38 \\ 
J0205+6449 & 78 & 88 \\ 
J2229+6114 & 55 & 46 \\ 
J1420-6048 & 30 & 35 \\
J2021+3651 & 75 & 85 \\
J1057-5226 & 75 & 69 \\ \hline
\multicolumn{3}{l}{}%
\end{tabular}
\caption{NOTES.-Col.(1):Pulsar name.Col.(2):The inclination angle.
Col.(3):The viewing angle.}
\end{table}

\begin{figure}\label{pulse}
\includegraphics[width=75mm]{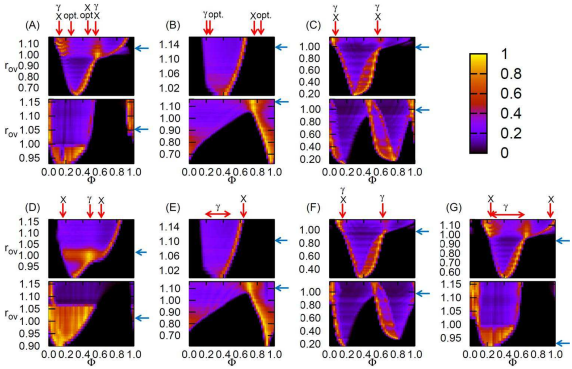}
\caption{The intensity maps for seven pulsars. Upper left to right: PSRs J0835-4510 (A), J0659+1414 (B) and J0205+6449 (C). Lower left to right: PSRs J2229+6114 (D), J1420-6048 (E), J2021+3651 (F) and J1057-5226 (G). In each sample, upper panel is outward emission and lower is inward emission. The blue horizontal arrows show best fit values of $r_{\rm ov}$ for $\gamma$-ray and  X-ray emission regions. The red vertical arrows show the phase of peaks. The red horizontal arrows in (E) and (G) show the phase range of broad peaks. See Kisaka \& Kojima (2011) \citep{KK11}.}
\end{figure}

It is worthwhile to explore the general dependence of the altitude $r_{\rm ov}$ on other characteristics if any, although there may not be enough data for a proper statistical analysis. In Fig. 2, $r_{\rm ov}$ is plotted as a function of inclination angle $\alpha$, spin-down luminosity $L_{\rm sd}$, characteristic age $\tau_{\rm c}$ and surface dipole magnetic field $B_{\rm s}$. We found that there is a significant correlation between $r_{\rm ov}$ and the inclination angle $\alpha$ only; the relations of $r_{\rm ov}$ with the other parameters are very weak. This correlation suggests that the deviation from a vacuum rotating dipole field is large for small inclination angle. It is very interesting to compare this result with that in a force-free magnetosphere. Bai \& Spitkovsky (2010b) \citep{BS10b} proposed that the separatrix layer at an altitude of 0.90-0.95 times the height of the last-open field line is relevant to emissions in a three-dimensional inclined force-free magnetosphere. This altitude, which is not exactly symmetric with respect to the magnetic azimuthal angle $\phi_{\rm m}$, but can be approximated by the value at $\phi_{\rm m}=0$, is plotted in Fig. 2 as purple downward and blue upward triangles. Two linear fitting lines are also shown. The altitude $r_{\rm ov}$ decreases with the inclination angle $\alpha$ in both our model and the separatrix layer model of a force-free magnetosphere.
\begin{figure}\label{r-a}
\includegraphics[width=55mm, angle=270]{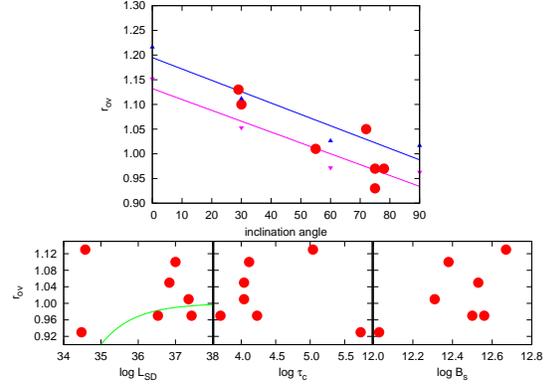}
\caption{The relation between $r_{\rm ov}$ and inclination angle (upper), spin-down luminosity (lower left), characteristic age (lower middle) and surface magnetic field (lower right). The altitudes corresponding to the separatrix layer model are shown as purple downward and blue upward triangles in the upper panel. The two lines are linear fitting lines for the separatrix layer model. The light green curve in the lower left panel shows the relation $(1-r_{\rm ov})=(10^{33}{\rm erg}\ {\rm s}^{-1}/L_{\rm sd})^{1/2}$. See Kisaka \& Kojima (2011) \citep{KK11}.}
\end{figure}

The thickness of the gap region, $w$, is not known, but it is sometimes assumed to decrease with the spin-down luminosity $ L_{\rm sd}$ \citep{Wa09, RW10}. We have $w=1-r_{\rm ov}$, if the lower boundary of the gap is fixed as the last-open field line in the vacuum dipole field. This assumption is tested in the lower left panel of Fig. 2, in which the relation $(1-r_{\rm ov}) \approx (L_{\rm sd} / 10^{33}{\rm erg}\ {\rm s}^{-1})^{-1/2}$ is plotted as a light green curve. (The curve is not fitted to the data points.) This suggests that the assumption of maximum altitude, $r_{\rm ov}=1.0$, is not a good one. This discovery affects expected number of the $\gamma$-ray pulsars in the observation. From geometrical reason, the pulsed emission by caustics is limited to a certain range between inclination and viewing angles. 
\begin{figure}\label{a-z}
\includegraphics[width=45mm]{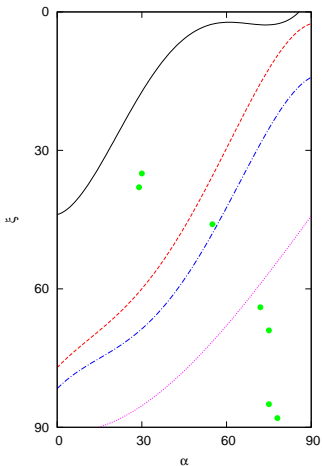}
\caption{The observable range for $\gamma$-ray pulsars in the $\alpha$-$\xi$ plane for the outer gap model. Black solid curve shows the boundary of observable pulsars using linear fitting line for the separatrix layer model. Red dashed, blue dash-dotted and purple dotted curves show the boundary with $r_{\rm ov}=$0.95, 0.90 and 0.70, respectively. Light-green curcles show the pulsars in Table 1. See Kisaka \& Kojima (2011) \citep{KK11}.}
\end{figure}

Romani \& Watters (2010) \citep{RW10} showed the range of observable pulsars with $r_{\rm ov}=$0.95, 0.90 and 0.70 for outer gap model in their Fig. 16. We recalculate it and show the result in Fig. 3. The observable range of viewing angle $\xi$ is below the curves. Our finding in Fig. 2 is that $r_{\rm ov}$ is a function of the inclination angle, which is similar to that of the separatrix layer model. We also show the observable range by the empirical relation obtained in Fig. 2 as black solid line, for which the altitude is chosen as 0.925 times the height of the last-open field line in force-free magnetosphere. The figure shows that sources with low inclination and viewing angles become observable. For example, pulsar with the inclination angle $\alpha =30^{\circ}$ can be detected for $\xi > 60^{\circ}$ for $r_{\rm ov}=$0.95, but for $\xi >30^{\circ}$. Thus expected number increases approximately twice for sources with the low inclination and viewing angles.

\section{SUMMARY}

We have calculated the light curves of emissions using the TCS08 outer gap model and compared them with observed multi-wavelength light curves. We find that the model can successfully explain the peak positions of multi-wavelength light curves. In order to determine the altitude of the emission region, the observed X-ray light curve is important, especially when there is a single peak in the $\gamma$-ray light curve.

The best-fit values of the altitude of the emission region for PSRs J0659+1414 and J1420-6048, suggest a deviation from the last-open field lines of a vacuum dipole field. The real last-open field lines lie inside those of vacuum dipole field, $r_{\rm ov}<1.0$. This shift suggests that the lower boundary is very similar to that of a force-free magnetosphere. We find that the altitude of the emission region is correlated with inclination angle. This relationship is also very similar to that in a force-free magnetosphere. The lower boundary of emission region has been assumed to $r_{\rm ov}=1$ so far, but our model fits do not support it. This modification of the boundary of the magnetosphere suggests that the pulsars with low inclination and viewing angles are likely to be detectable. Thus the expected number in the future observation in the previous works is underestimated for the sources with low inclination and viewing angles.

\bigskip 
\begin{acknowledgments}
The authors thank S. Shibata, J. Takata and T. Wada for much valuable 
discussion.
This work was supported in part by the Grant-in-Aid for Scientific 
Research from the Japan Society for Promotion of Science(S.K.) 
and from the Japanese Ministry of Education, Culture, Sports,
Science and Technology(Y.K.  No.21540271).
\end{acknowledgments}

\bigskip 

\end{document}